# Computational modelling of peritoneal dialysis: an overview

Sangita Swapnasrita[1, 2], Joost C de Vries[2], Carl M. Öberg[3], Aurélie MF Carlier[1*], Karin GF Gerritsen[2*]

[1] MERLN Institute for Regenerative Medicine, Maastricht University, Universiteitssingel 40, 6229 ER Maastricht, The Netherlands

[2] Department of Nephrology and Hypertension, University Medical Center Utrecht, Heidelberglaan 100, 3584 CX Utrecht, The Netherlands

[3] Department of Clinical Sciences Lund, Division of Nephrology, Skåne University Hospital, Lund, University, Lund, Sweden

*Both contributed equally

*Corresponding author*

*Email:* *K.G.F.Gerritsen@umcutrecht.nl*, a.carlier@maastrichtuniversity.nl

**Abstract**

Peritoneal dialysis (PD) is becoming more popular as a result of a rising interest in home dialysis, lower intrusion in social life and longer preservation of residual kidney function. However, PD has several important drawbacks: small solute clearance is relatively low compared to hemodialysis and technique survival is limited. Application of continuous flow, sorbent-based dialysate regeneration and novel glucose-sparing PD solutions are some solutions proposed to address the limitations of PD. To optimize and personalize current and novel PD therapies, patient peritoneal characteristics interacting with PD techniques need to be studied together and separately as they interplay. However, considering the multitude of parameters, it would be difficult, expensive, and time consuming to optimize all parameter settings only with the help of clinical trials. Mathematical modelling is an exciting tool to dissect these interacting processes and comprehend PD techniques better at a patient specific level. In this review, we look at the history of computational PD models, explore the many ways a computational PD model can be constructed and review the various existing PD models that can be used to optimize and personalize PD treatment.

## 1 PERITONEAL DIALYSIS

Over 850 million people worldwide, i.e. ~1 in every 10, suffer to some degree from chronic kidney disease [1] among which over 3.8 million are on dialysis, either peritoneal dialysis (PD) or hemodialysis (HD) [2]. During PD dialysis fluid is instilled into the abdominal cavity via a permanent catheter. The lining of the abdominal cavity (the peritoneum) acts as a semipermeable membrane for solute and water transport. PD removes waste products from blood plasma by diffusion and convection, and excess water by osmosis across the peritoneal membrane into the dialysis fluid in the abdominal cavity (see Figure 1). Continuous ambulatory PD (CAPD) and automated PD (APD) are the two main modalities of PD used in routine practice. CAPD is a continuous 24 hour dialysis therapy where the dialysis fluid (1-2.5L) is exchanged (drained and instilled) 3-4 times daily manually through a catheter [3]. A typical dwell lasts between 4-8 hours. APD on the other hand uses a cycler (with patient specific treatment parameters) connected to the abdominal catheter throughout the session (preferably at night) to automatically perform the dialysis fluid exchanges over a period of time (usually 4-5 per night) [4]. This is often combined with a long dwell (14-15h) during the day. Both CAPD and APD are usually conducted every day of the week.

PD has several advantages as compared to HD: it allows for continuous gradual removal of waste (instead of intermittent HD, which is characterized by a 'saw-tooth pattern'), does not require blood access, provides more patient autonomy as the treatment is performed at home, and is less expensive. Residual kidney function is also better preserved [5] compared to HD. However, PD has important shortcomings. Technique survival is limited (median 3.7 years [6]) due to recurrent peritonitis (inflammation of the peritoneal membrane), catheter dysfunction or membrane failure (due to exposure to high (harmful) dialysate glucose concentrations required for osmotic water removal), and small solute clearance is relatively low. Due to technique failure or low small solute clearance (with disappearance of residual diuresis) patients often have to switch to HD after several years [6, 7].

Despite 60 years of progress in PD, it is still only used by ~11% of dialysis patients with considerable variations across countries, mostly due to non-medical reasons [8-10]. In comparison, Hong Kong has ~80% end stage kidney disease (ESKD) patients on PD to reduce expenditure [11] after an implementation of "PD first" policies. Countries like Mexico, Thailand and Singapore also have high uptake of PD due to availability of medical personnel to assist with PD. To increase PD usage and better personalisation of PD regimens, there are several novel techniques under development or already introduced to the market such as automated PD [12, 13], continuous flow PD [14, 15], sorbent assisted PD [16-18] along with many other sub-varieties (Figure 1). Tidal PD is a form of automated PD which exchanges only portions of the dialysis fluid in shorter cycles (after an initial complete fill) so that the late 'slow flow' segment of the drain phase is kept to a minimum [19], to reduce abdominal pain [20]. In continuous flow PD (CFPD), there is a continuous flow of dialysate through an inflow catheter and outflow catheter [14, 15, 21]. The continuous flow of fresh dialysate through the abdominal cavity maintains a large plasma–dialysate concentration gradient, increasing solute transfer across the peritoneal membrane. Dialysate glucose concentration can be kept nearly constant, thereby maintaining a constant osmotic gradient and ultrafiltration rate at lower (peak) glucose concentrations than in conventional PD [22], possibly slowing functional deterioration of the peritoneal membrane and reducing peritonitis rate (associated with high glucose concentrations). For CFPD, two single lumen catheters or a double lumen catheter may be applied. Sorbent-assisted PD regenerates the dialysis fluid using a sorbent cartridge [16, 23, 24]. The sorbent-assisted PD systems that are currently under development employ rapid cycling of dialysis fluid via a single lumen catheter. With further technical improvements (*e.g.* online production of peritoneal dialysate, novel PD solutions with improved biocompatibility and/or glucose-sparing), PD treatment may improve further.

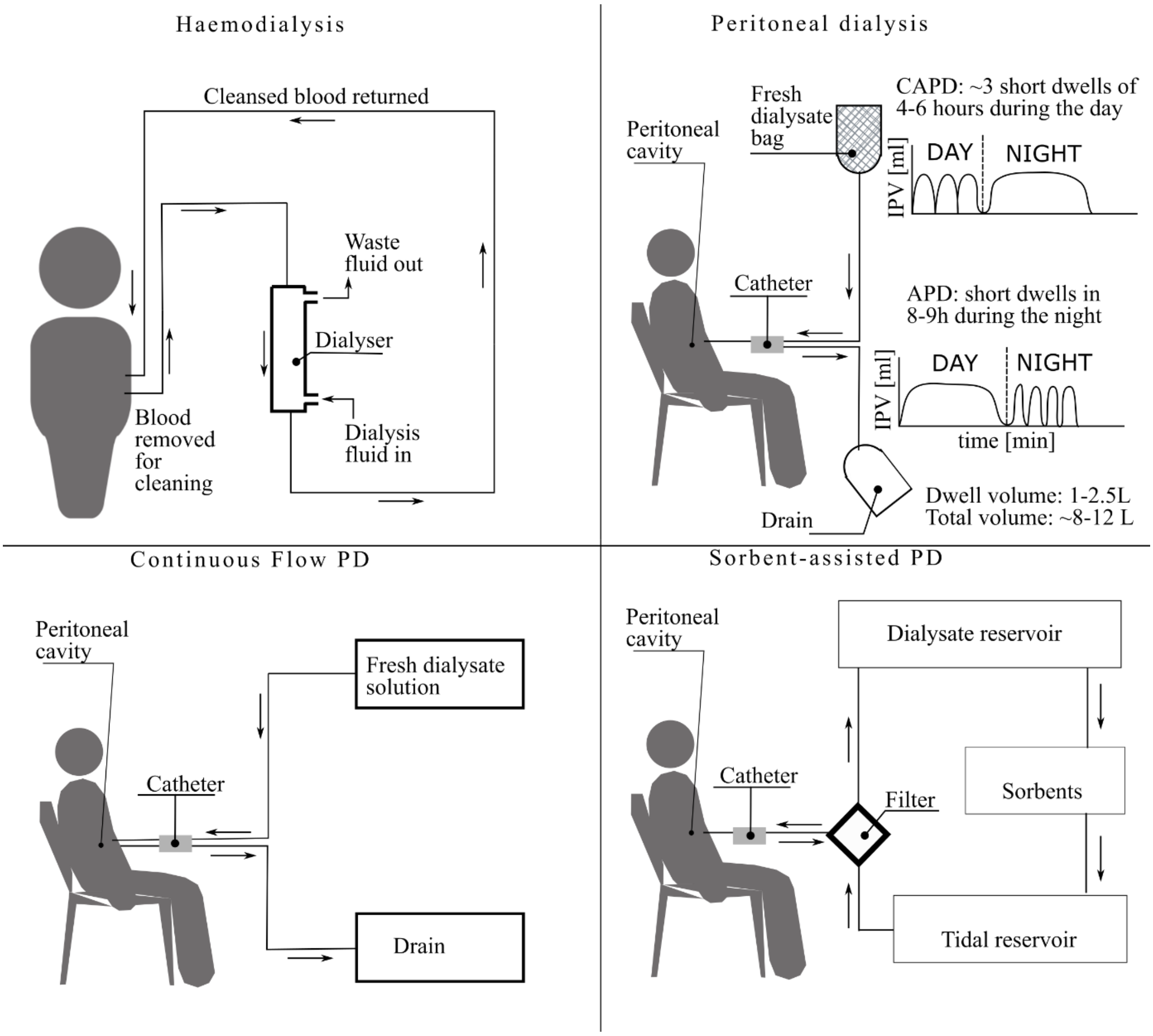

**Figure 1:** Different modes of dialysis therapy techniques available for the chronic kidney patient. IPV = Intraperitoneal volume.

The development of new technologies often takes a considerable amount of time owing to the learning curve, ensuring transparency, ethical requirements, trial regulations *et cetera* [25]. The new (post-2020) European Union Medical Device Regulation (EU-MDR) rules related to manufacturing of medical devices are more stringent, also leading to delays in certification. As such, there is a need to supplement the pre-clinical research and clinical trials with other methodologies to accelerate the design, manufacturing and marketing of novel PD technologies. Computational models are a revolutionary tool in the field of health, medicine and life sciences due to the ease of optimisation, non-intrusiveness and most importantly, the interpretation of complex interdependent physical processes. It is being increasingly used to study the influence of biological processes in medical devices, including for example thrombogenic reaction to biomaterials [26], controlled drug release [27], effect of implant surface roughness on protein adsorption [28]. Moreover, it is often an inexpensive and effective way to simulate complex natural phenomena. In this review, we focus on computational models for PD. Hereto we look at the history of PD modelling to highlight the advances and remaining lacunas as well as learn which level of simplicity and physiological detail is necessary to reach a particular research aim. This review of general concepts of PD models is specifically intended for experimental scientists or clinicians with little or no computational modeling experience, but who are interested in introducing modeling into their research practice. By providing an overview of existing computational models, the individual aspects of the models, and the scope of the model, we aim to simplify, enhance and accelerate the integration of modeling into clinical practice to promote better understanding of device-patient interaction. In section 6, we show an example of a scenario to illustrate how one can setup a mathematical model of PD using existing models highlighted in this review.

## 2 COMPARTMENTAL MODELS FOR PERITONEAL DIALYSIS

There are many types of computational models, including linear and non-linear, deterministic and stochastic, discrete and continuous, spatial and non-spatial ones. For general reviews on computational modeling we refer to Yates *et al.* [29], King *et al.* [30] and Brown *et al.* [31]. The most common design for models of PD is the compartmental model. They are simple in nature for physiological kinetic and dynamic modelling. In compartmental models, the body is divided into theoretical compartments such as the peritoneal membrane, peritoneal cavity and total body water (examples in Figure 2). The general assumption is that a tissue or organ can be represented as a homogeneous compartment, governed by conservation of mass and other properties of interest (e.g. charge). A compartmental model consists of volumes connected by fluxes of some entity. Each compartment can be a volume representation (in case of PD) and it characterises the essential physics and chemistry of the biological environment. The flow rates and interactions between the compartments are described by the parameters of the compartmental model. Some PD models use a single compartment model with just the peritoneal cavity and body considered to be a constant source of solutes [32] while some models include multiple compartments (*e.g.* the distributed model, [33]), with the peritoneal tissue, peritoneal cavity and interstitium as different compartments. Compartment models are lumped models but despite their simplicity, they usually capture the underlying physical and biological phenomena well, displaying why they are commonly used for PD modelling.

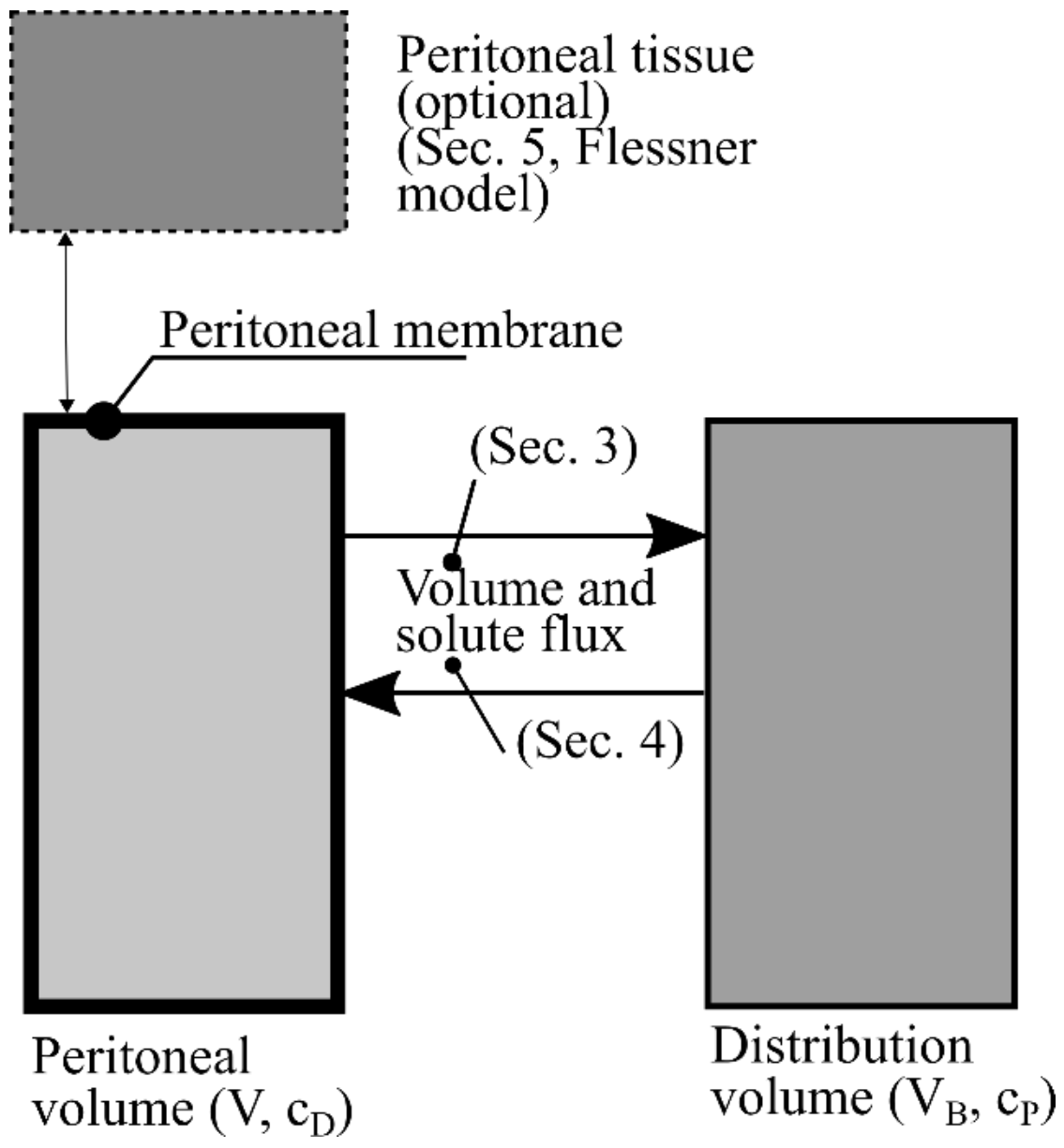

**Figure 2:** Compartmental PD modelling showing compartments (optional compartments in dotted lines), membranes and fluxes. $V$ = peritoneal compartment volume, $V_B$ = Body compartment volume, $c_D$ and $c_B$ = peritoneal and body solute concentration.

The models usually employ fluid and solute mass balance equations, which are described in section 3 and 4. It depends on the type of fluid flow (static or continuous), complexity (single compartment or distributed) and membrane model (continuous semi-permeable membrane or 3 pore) to decide which and how the fluid and solute flows should be properly modelled. In case of PD, physiologically the barrier is a semi-permeable peritoneal membrane. The compartments can exchange the fluid and solutes without any barrier or with a barrier.

The simplest interpretation of the peritoneal membrane is a simple membrane that allows passive diffusion of solutes [34-36]. The diffusive flow of molecules ($J_s$ in mol/s) will be dependent on the concentration gradient $\frac{\mathrm{d}c}{\mathrm{d}x}$ across the barrier:

$$J_s \propto \frac{\mathrm{d}c}{\mathrm{d}x}$$

$$J_s = -DA\frac{\mathrm{d}c}{\mathrm{d}x}$$

where $D$ is the diffusion coefficient ($\mathrm{m}^2\ \mathrm{s}^{-1}$), $A$ is the effective surface area available for diffusion ($\mathrm{m}^2$), $c$ is the concentration (mol/l) and $x$ is the distance across the membrane (m). The negative sign is because the solute flux is in the opposite direction of the concentration gradient. Integrating the above equation from $x$=0 to $x$=Δ$x$ (across the membrane),

$$J_s = -PS \cdot (c_2 - c_1) \qquad 2.1$$

where $PS = DA/\Delta x$, and $c_1, c_2$ are the concentrations on either side of the membrane, at $x$=0 and $x$=Δ$x$. Equation 2.1 can be used to calculate the diffusive plasma clearance of a particular solute from $c_1$ to $c_2$ ($J_s/c_1$) or vice versa ($J_s/c_2$). The factor $PS$ is known as the diffusion capacity of the solute species and is defined as the maximal absolute diffusive plasma clearance (when the concentration is zero on one side of the membrane). Many abbreviations and terms are in use for this parameter such as mass transfer area coefficient (MTAC), $k_{BD}$, *et cetera*.

## 2.1 THREE-PORE MODEL OF PERITONEAL DIALYSIS

A perplexing thing about the peritoneal membrane is that it is a semi-permeable membrane that allows passage of albumin and other large proteins to a limited extent but also restricts the bulk movement of electrolytes [16, 37-40]. In 1981, Nolph *et al.* stated, “It is a system that displays characteristics of some very large-pore radii when assessed by diffusion studies, and some very small-pore radii when assessed by ultrafiltration and solute reflection coefficients” [41]. The reflection coefficient, $\sigma$ describes the convective hindrance of a molecule (a value of 100% means convective clearance does not occur). They were the first to hypothesize that the peritoneal membrane is a heteroporous membrane with both small and large pores. Small pores with low solute permeability would be located in the proximal capillary and facilitate ultrafiltration due to the relatively high hydraulic pressure and high osmotic gradient in the proximal capillary. Towards the venous capillaries the hydraulic pressure drops and the oncotic pressure increases as a result of increased protein concentration caused by capillary ultrafiltration. A predominance of large "pores" in the venular capillaries with high permeability would facilitate diffusive solute exchange. Glucose would be more readily absorbed resulting in lower osmotic pressure and the ultrafiltration rate would be reduced at the venular capillaries also because of the relatively low hydrostatic and high oncotic pressure. In summary, most ultrafiltration would occur through the “small” pores with low hydraulic permeability and most solute exchange through highly permeable “large” pores (Figure 3 A). Aside from these two pathways, there is also a separate channel for movement of water that is inaccessible to solutes. To account for all three pathways, the three pore model was developed.

In 1991 Rippe *et al.* [42] proposed the three pore model (Figure 3 B). It is the most common representation of the peritoneal membrane. It divides the body into two compartments – the distribution volume (specific to each solute) and the peritoneal cavity. The solute and volume flows through the different pores are defined by the Starling and Patlak equations respectively (explained in equation 3.1 and 4.1). The most abundant (99% of the pore fraction) is the protein restrictive water-soluble pathway (15 to 36 Å) responsible for 90% of the hydraulic conductance. The “large pores” of 250 Å constitute 0.01% of the pore population and represent 8% of the hydraulic conductance and the final pore fraction is the ultrasmall pores (0.99%) and responsible for only 2% of the hydraulic conductance. The “small pores” correspond to the gaps between the endothelial cells and the “large pores” correspond to the venular interendothelial pathways. These ultrasmall pores (2.3 to 15 Å) are only permeable to water (reflection coefficient is unity for all solutes) and were later shown to represent the aquaporin (AQP-1) water channels firstly identified only a year after Rippe proposed their existence [43]. The Rippe model can predict the transport of not only small and intermediate solutes but also large solutes with reasonable accuracy.

Venturoli *et al.* proposed a series of two porous membranes (Figure 3 C) to model the bi-directional clearances of macromolecules [44]. The first layer is the three-pore membrane by Rippe, identified as the capillary endothelium. This is followed by a second three pore membrane consisting largely of large pores (95%) and 2 and 3% respectively of transcellular and small pores, which is a lumped representation of extracellular interstitium. This model is able to mathematically explain the build-up of tracer albumin [45] seen in rats which the one layer three pore model is not able to.

Another approach to modelling the peritoneal membrane is the distributed model by Flessner [33] (Figure 3 D). They model the peritoneal tissue space as a tube-and-shell exchanger with a constant void space (blood capillary or the plasma space) and the interstitium. Water movement occurs throughout the tissue while solute convection occurs only across the plasma capillaries. All plasma capillaries are the same shape and size, which makes this a single pore model. This model is also complementary to the Venturoli model [44] in that it allows to calculate the accumulation and release of certain substances dissolved in the dialysis fluid. Because of the additional modelling of the surrounding tissue space, the Flessner model is able to give insight into the mechanisms ongoing in the peritoneal tissue which could, for example, potentially be useful for modelling drug transport in patients being administered intra-peritoneally.

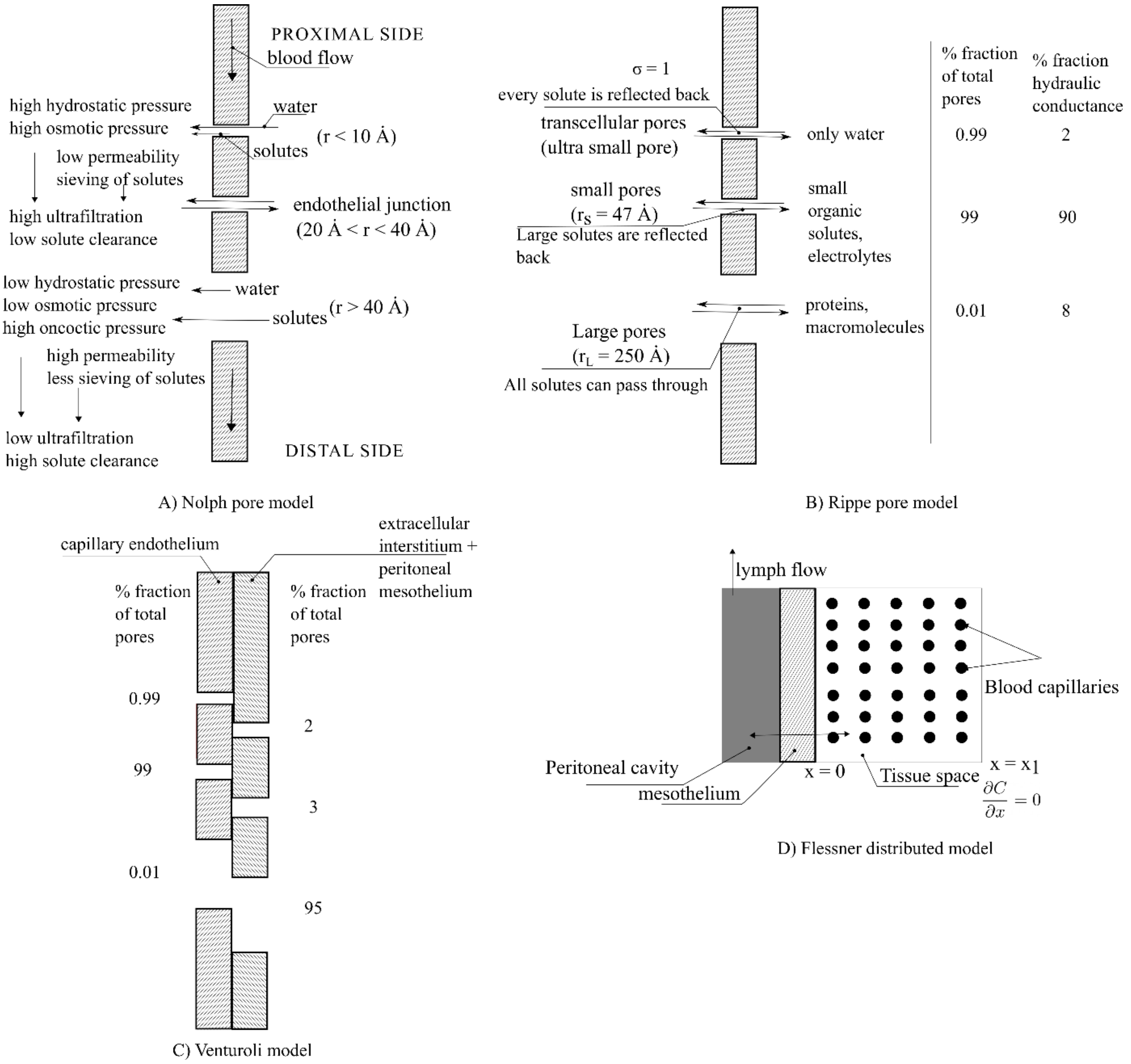

**Figure 3:** A) Nolph three-pore model (high permeable and low permeable pores), B) Rippe three-pore model (distinction based on size, population and contribution to ultrafiltration), C) Venturoli model (two porous membrane side-by-side), and D) Flessner distributed model (tissue space with uniformly distributed capillaries- essentially a “single pore model” for the capillaries).

# 3 VOLUME FLOW CALCULATIONS

The osmotic gradient created by a hypertonic dialysis fluid drives the flow of water from the plasma by osmosis, commonly referred to as ultrafiltration [46]. In APD, underestimation or overestimation of the ultrafiltration volume may result in setting wrong drain volume of the peritoneal cavity. The use of wrong drain settings may cause discomfort to the patient and reduce the efficiency of the dialysis session. As such, it is important to properly calculate all the flows occurring during PD, *i.e.* the amount of net ultrafiltrate is determined by

- the glucose concentration used in the dialysis fluid and subsequent water flow due to the osmotic gradient
- the flow of dialysate into the peritoneal cavity at a certain flow rate (APD, CFPD) or by gravity (static dwell, CAPD)
- the lymphatic flow of fluid from the intraperitoneal space towards the lymphatic space
- other physical (and patient-specific) attributes such as the condition of the peritoneal membrane, intraperitoneal volume and pressure, lymphatic system of the patient etc.

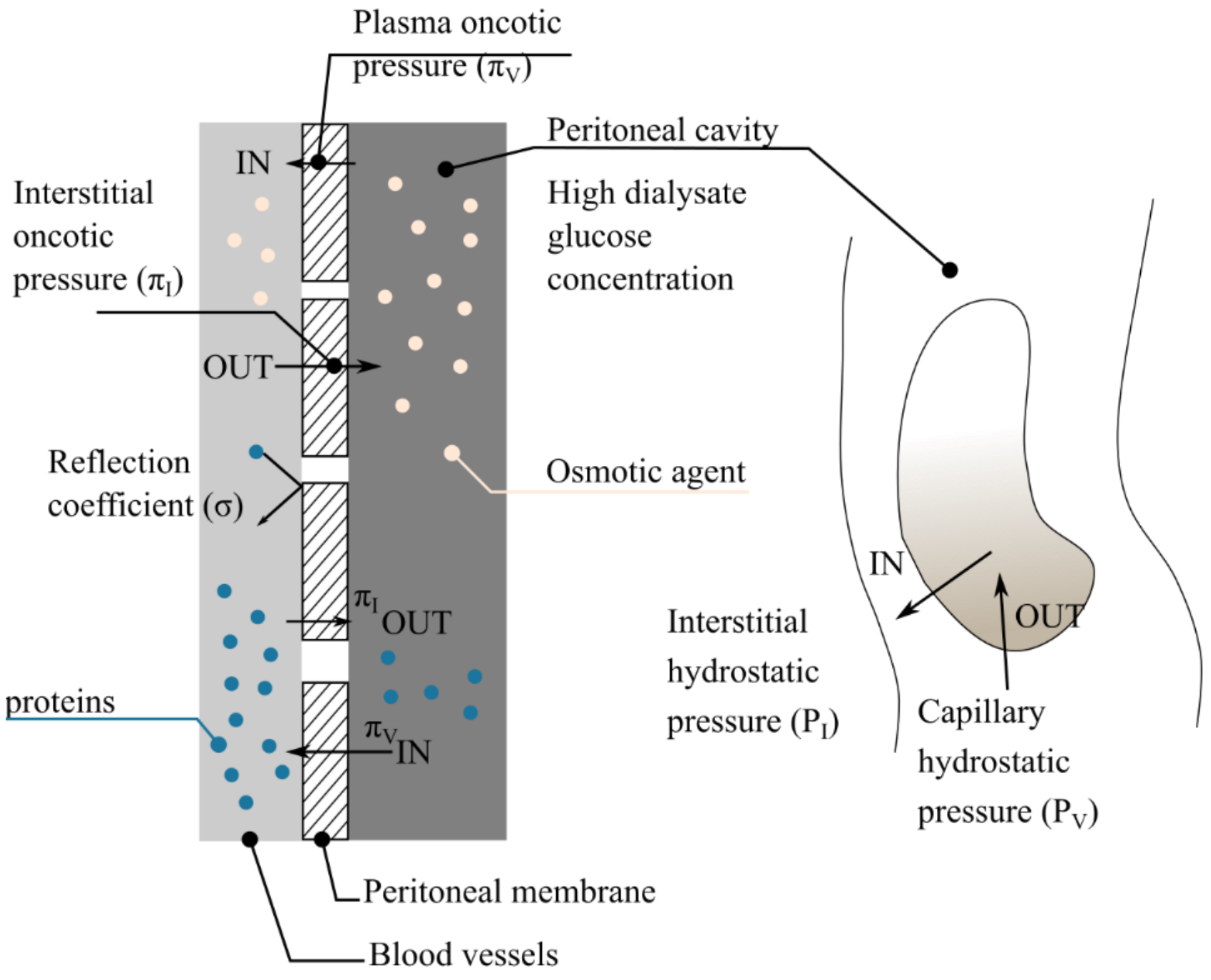

**Figure 4:** Fluid flow and solute movement across the peritoneal membrane due to hydrostatic pressures and osmotic gradient (see text). A high concentration glucose solution (yellow dots) is used which means glucose gradient is in a different direction at the beginning of the dwell compared to that of other solutes and proteins (blue dots), which are usually in equilibrium or lower in the dialysate. The striped boxes represent the peritoneal membrane with three types of pores (section 2.1). I denotes interstitial parameters and V denotes the plasma parameters. IN = from dialysate to blood and OUT = from blood to dialysate.

There have been limited efforts to model volume change in the intraperitoneal cavity. We describe the different ways of estimating peritoneal volume change and the ways to obtain the parameters for the flux equations in further subsections.

To describe the water flow (volume flow), the Starling equation is often used, which is essentially a modified version of Ohms law, $\text{Flow} = \text{Hydraulic conductance} \cdot \Delta\text{Pressure}$. The original Starling equation was written to describe the fluid flow from a capillary to the interstitial space [47]. This equation was later corroborated by Kedem and Katchalsky from a thermodynamics point of view [48]. As fluid builds up in the peritoneal cavity, it exerts an outward interstitial force on the peritoneal membrane ($P_{\mathrm{I}}$, Figure 4). Due to the fluid moving into the interstitial space, the corresponding fluid pressure rises and this opposing pressure is the capillary hydrostatic pressure ($P_{\mathrm{V}}$, Figure 4). The plasma proteins exert a colloid osmotic pressure or oncotic pressure ($\pi_{\mathrm{V}}$, Figure 4) to draw water back from the peritoneal cavity to the plasma (opposite to glucose). The osmotic agent in the dialysate exerts an outward osmotic pressure ($\pi_{\mathrm{I}}$, Figure 4). To model the above, equation 3.1 consists of two inward forces, i.e. interstitial hydrostatic ($P_{\mathrm{I}}$) and plasma colloid osmotic pressure ($\pi_{\mathrm{V}}$) and two outward forces, i.e. hydrostatic pressure ($P_{\mathrm{V}}$) and interstitial osmotic pressure ($\pi_{\mathrm{I}}$) (see Figure 4). Here, we will also take into account that the peritoneal membrane is semi-permeable and will have different reflection coefficients for solutes of different size. Note that by inward or outward, we mean into or out of the peritoneal cavity. We refer the reader to the appendix for an overview of all notations.

The volume flow (ml $\text{min}^{-1}$) across the peritoneal membrane is thus modelled as follows:

$$\frac{dV}{dt} = J_v = L_pS[(P_I - P_V) - \sigma(\pi_I - \pi_V)]$$
$$= L_pS\left[\overbrace{\Delta P}^{hydrostatic\ pressure\ diff} - \sigma \underbrace{\Delta\pi}_{osmotic\ pressure\ diff}\right] \quad 3.1$$

The volume in the peritoneal cavity is $V$, the hydraulic conductance of the peritoneal membrane is given by $L_pS$ ($\text{l}\,\text{s}^{-1}\text{mmHg}^{-1}$) and the reflection coefficient of the solute is given by $\sigma$ (dimensionless). I denotes interstitial parameters and V denotes the plasma parameters. We have to make certain modifications to equation 3.1 as the fluid is not in full contact with the peritoneal membrane, which means that the whole membrane surface cannot be taken into account for the calculation of the volume flow but rather the peritoneal surface area in contact with the dialysis fluid ($af$, fraction of the peritoneal surface area in use, dimensionless). This gives us modified equation 3.2.

$$J_\text{v} = L_pS[\Delta P - \sigma\Delta\pi]af \quad 3.2$$

In addition to the fractional peritoneal surface area in use, we can divide the volume flow among the various types of pores present in the peritoneal membrane. If we use the classical 3-pore model here, we can divide the net volume flow into flows across the ultrasmall, small and large pores by multiplying their contribution to the ultrafiltration ($\alpha$, dimensionless).

$$J_\text{vc} = L_pS[\Delta P - \sigma\Delta\pi] \cdot af \cdot \alpha_\text{C}$$
$$J_\text{vS} = L_pS[\Delta P - \sigma\Delta\pi] \cdot af \cdot \alpha_\text{S} \quad 3.3$$

$$J_{v\mathrm{L}} = L_p S[\Delta P - \sigma \Delta \pi] \cdot af \cdot \alpha_{\mathrm{L}}$$

where $J_{\mathrm{vC}}$,$J_{\mathrm{vS}}$ and $J_{\mathrm{vL}}$ are the volume flows across the ultrasmall, small and large pores respectively.

The change in volume is then calculated as a total of all the volume flows and the net lymphatic flow [32],

$$\frac{dV}{dt} = J_{vC} + J_{vS} + J_{vL} - L \qquad 3.4$$

where $L$ is the lymphatic flow rate (l $\mathrm{s}^{-1}$) (a sum of all lymphatic flows that drain the peritoneal cavity, i.e interstitial, diaphragmatic, pelvic and omental [49]), considered to be around 0.15-0.5 ml/min (0.216-0.72 L/day).

Öberg *et al.* extended equation 3.4 for APD [50] and CFPD [32],

$$\frac{dV}{dt} = J_{vC} + J_{vS} + J_{vL} - L + J_{fill} - J_{drain} \qquad 3.5$$

where $J_{\mathrm{fill}}$ and $J_{\mathrm{drain}}$ are the fill and drain flow rates during CFPD (l $\mathrm{s}^{-1}$).

The initial fill volume and drain volume are in general known. The residual volume, $V_{\mathrm{r}}$ may be estimated from the dilution of solute concentration (e.g. albumin, creatinine or total protein) measured in the drained effluent, by measuring concentration just after instillation of a known fill volume $V_{\mathrm{fill}}$ [51] (equation 3.6):

$$V_r = \frac{V_{fill} * c_0}{(c_{drain} - c_0)} \qquad 3.6$$

where $c_0$ is the measured concentration of the solute just after filling the peritoneal cavity (rapid mixing is assumed) and $c_{\mathrm{drain}}$ is the concentration of the solute in the drain bag collected in the previous session. Knowing these volumes the net ultrafiltration volume, UF, can be calculated as,

$$(Net)UF = V_{drain} + V_{r,t2} - V_{fill} - V_{r,t1} \qquad 3.7$$

where $t1$ is the time of instillment of the dialysis fluid and $t2$ is the time of draining.

Lymphatic flow may or may not be considered depending on the patient and the model.

Volume flows are also estimated assuming that the rate of ultrafiltration is known or can be calculated [52].

## 3.1. HOW TO ESTIMATE VOLUME FLOW PARAMETERS?

Few parameters are derived from experimental observations and some are obtained from optimisation of models.

### 3.1.1 Difference in hydrostatic pressure, $\Delta P$

The hydrostatic pressure difference between the interstitium and the capillaries in most organs is usually around 10-17 mmHg (13.6-23.1 cm $H_2O$) [53]. Durand et al. measured the mean intraperitoneal hydrostatic pressure (IPP) to be 13 cm $H_2O$ [54]. However, in the case of a static PD dwell, the pressure builds up inside the peritoneal cavity due to ultrafiltration such that $IPP$ is a function of time for a patient in a sitting position. Twardowski et al. derived the empirical formula for dependence of IPP as a function of intra-peritoneal volume as follows [55],

$$\Delta P = \Delta P_0 + \frac{V_t - (V_{fill} + V_r)}{490} \qquad 3.8$$

where $P_0$ is the baseline hydrostatic pressure, for example 13 cm $H_2O$, $V_{\mathrm{fill}}, V_{\mathrm{r}}, V_t$ are the initial fill volume (1.0-2.5 l depending on the patient), residual volume (usually about 0.25 l) and the intraperitoneal volume at time $t$, respectively. $V_{\mathrm{fill}}$ is known and the residual volume $V_{\mathrm{r}}$ is calculated from equation 3.6. The peritoneal volume at different times, $V_{\mathrm{t}}$, can, for example, be calculated from the dilution of RISA (radioactive $^{125}$I serum albumin) [56, 57] during the dwell or from direct volume recovery technique [58].

For continuous flow PD, the hydrostatic pressure changes depending on the drain and fill flow rate, and will generally increase or decrease with the momentaneous intra-peritoneal volume.

### 3.1.2 Reflection coefficients, $\sigma$

Sieving or reflection coefficients for solutes can be determined from experiments [59, 60] or from analytical equations such as that of Drake et al. [61]

$$\sigma = \frac{16}{3}\lambda^2 - \frac{20}{3}\lambda^3 + \frac{7}{3}\lambda^4 \qquad 3.9$$

where $\lambda$ (dimensionless) is the ratio of the solute radius to the membrane pore radius. Thus, the reflection coefficients are calculated separately for different pore sizes and solutes.

Other analytic solution widely used for three pore model especially is [62],

$$\sigma = 1 - \frac{\left(1 - \frac{\lambda}{3}\right)(1-\lambda)^2[2-(1-\lambda)^2]}{\left(1 - \frac{\lambda}{3} + \frac{2}{3}\lambda^2\right)} \qquad 3.10$$

Theoretically, one can also obtain the (steric) reflection coefficient from the equation by Anderson et al. if one knows the equilibrium concentration of the solute on either side of the peritoneum, i.e., the blood plasma and the peritoneal cavity [63].

$$\sigma = (1 - \Phi)^2 \qquad 3.11$$

where $\Phi$ is the equilibrium partition coefficient (dimensionless), given by $\Phi = \left(\frac{C_1}{C_2}\right)_{\mathrm{eqlb}}$ and $c_1$ and $c_2$ are equilibrium concentrations on either side of the peritoneal membrane.

### 3.1.3 Osmotic conductance to glucose

The osmotic conductance to glucose of the peritoneal membrane is usually calculated from double mini-peritoneal equilibration test (dm-PET) described by La Milia *et al.* [64]. In the clinic, a 60 min dwell 1.5% (~83 mmol/l) glucose is followed by 60 min 4.25% glucose (~236 mmol/l) and the difference between the drained volume is used to calculate the osmotic conductance to glucose (OCG or $\sigma_{\mathrm{g}} L_p S, \frac{\mathrm{\mu L}}{\mathrm{min}}/\mathrm{mmHg}$). Assuming that the initial volume flow is due to osmosis only, from equation 3.2, we get

$$OCG = \frac{\Delta J_{\mathrm{v}}}{\Delta\pi_{4.25} - \Delta\pi_{1.5}} = \frac{\Delta J_{\mathrm{v}}}{RT(c_{4.25} - c_{1.5})}$$

$$OCG \approx \frac{V_{4.25} - V_{1.5}}{100} \qquad 3.12$$

$R$ is the gas constant ($0.082\ \mathrm{litre \cdot atm\ K^{-1}\ mol^{-1}}$) and $T$ is the absolute temperature (K). Typical values for OCG are between 3-4 µL/min per mmHg. Knowing the reflection coefficient of glucose from equation 3.9 or 3.10 or from experiments, $L_p S$ can be calculated to be around 0.08 ml/min per mmHg [42, 65-67]. OCG may also be derived from a single dwell as described in the article by Martus et al. [53].

### 3.1.4 Fractional peritoneal surface area in contact with the dialysate, $af$

Keshaviah *et al.* calculated the empirical relationship between fill volume and peritoneal surface area in contact [68, 69],

$$af = \frac{16.18(1 - e^{-0.00077.V})}{13.3187} \quad 3.13$$

The fractional peritoneal surface area ($af$, dimensionless) in contact with the dialysate can also be obtained from stereological experiments by superimposing a grid over the CT scan of the peritoneum [70, 71] or by magnetic resonance imaging [72]. The fractional surface area in contact is an important determinant of the mass transfer coefficient of solutes. It can be improved by increasing the fill volume [73], agitating the dialysate or adding surfactants [74] but the latter has not been tried in humans. Equation 3.13 has been shown to be identical to the cube-square law for intra-peritoneal volumes up to ~2,300 ml [32].

### 3.1.5 Contribution to hydraulic conductance, $\alpha$

Fractional hydraulic conductances ($\alpha$) may be estimated by fitting the experimental results to the model [44, 69]. The usual values used in the three-pore model are given in Figure 3 B.

### 3.1.6 Ultrafiltration rate, $UFR$

There have been multiple ways of determining ultrafiltration rate ($\mathrm{l\,s^{-1}}$) throughout the literature. The average ultrafiltration rate may be estimated from the drain, $V_{\mathrm{drain}}$ and fill volume, $V_{\mathrm{fill}}$ during a static dwell [52],

$$UFR = \frac{V_{drain} - V_{fill}}{t} \quad 3.14$$

The UF rate can also be theoretically calculated [75, 76],

$$UFR = L_p S(\Delta P - \sigma \Delta \pi) - L \quad 3.15$$

Lymphatic flow is often considered to be constant throughout the dwell.

The UF rate is sometimes calculated as a function of time from the interpolated intraperitoneal volumes [77-82], as the UF rate usually decreases during a static dwell [75, 83, 84].

$$UFR = \frac{V_{t+\Delta t} - V_t}{\Delta t} + L \qquad 3.16$$

A simple empirical exponential model was used by Randerson et al. to capture the time dependent UF rate [85],

$$UFR = A(1 - e^{-\beta t}) \qquad 3.17$$

where $A$ is a fitting constant and $\beta$ is the time constant (a value of 0.0192 $min^{-1}$ was used).

Total ultrafiltrate volume can also be calculated from the dilution of initial dialysate albumin concentration, as done by Krediet *et al.* [86],

$$(Total)UF = \underbrace{\frac{c_{D,0}}{c_{D,t}} V_0 - V_0}_{net\ transcapillary\ UF} - \left( \underbrace{\frac{c_{D,0}}{c_{D,g}} V_0 - \frac{c_{D,t}}{c_{D,g}} V_t}_{lymphatic\ absorption} \right) \qquad 3.18$$

where $c_{D,0}$ and $c_{D,t}$ are the dialysate albumin concentrations at time 0 and time $t$, which is the end of the dwell, $c_{D,g}$ is the geometric mean of the dialysate albumin concentration, $\sqrt{c_{\mathrm{D},0} * c_{\mathrm{D},t}}$ and $V_0$ and $V_t$ are the intraperitoneal volumes at time 0 and $t$.

For CFPD, Öberg *et al.* theoretically derived the following relation for UF rate [32],

$$UFR = \frac{\sqrt{(J_{fill} + MTAC_{glu})^{\,2} + 4J_{fill}U_{max}} - (J_{fill} + MTAC_{glu})}{2} \qquad 3.19$$

where $J_{\mathrm{fill}}$ is the fill volume ($\mathrm{l\,s^{-1}}$), $MTAC_{\mathrm{glu}}$ is the diffusion capacity ($\mathrm{l\,s^{-1}}$) and $U_{\mathrm{max}}$ is a function of glucose concentration ($\mathrm{l\,s^{-1}}$). The equation slightly overestimates the ultrafiltration.

$$U_{max} = RT \cdot OCG \cdot c_{glu} - 3.1 \qquad 3.20$$

where $RT$ is the product of gas constant and the temperature in degree Kelvin $\left(\frac{\mathrm{mmHg}}{\frac{\mathrm{mmol}}{\mathrm{L}}}\right)$, OCG is the osmotic conductance to glucose $\left(\frac{\mu\mathrm{L}}{\mathrm{min}}/\mathrm{mmHg}\right)$ and 3.1 is a constant to account for the lymphatics, opposing forces and hydrostatic pressure gradient for other solutes in plasma.

Gotch also provided an empirical formula for CAPD using a dextrose solution [87],

$$(Total)UF = (184 + 512\ln(\%D))\ (1 - \exp(-0.02t)) \qquad 3.21$$

where $\%D$ is the w/w dextrose solution (dimensionless) and $t$ is the time elapsed. For other dialysate solutions similar curves for total UF *versus* time at different concentrations can be drawn to derive an analytical equation. The equation can then be used to determine the dialysate solution for the desired ultrafiltration.

Depending on the importance of ultrafiltration in the objective of the modelling efforts, one can opt for the simplified UF values such as equation 3.14 and 3.16 or for patient specific efforts equation 3.17 and 3.21. Equation 3.18 or 3.19 can provide UF estimates with sufficient precision but require measurement the concentration of albumin at different time-points, or the estimation of OCG and diffusion capacity of glucose, respectively.

# 4 SOLUTE FLOW CALCULATIONS

Solute flow across a semipermeable peritoneal membrane occurs because of two simultaneous processes. The first is diffusion due to the (electro)chemical gradient between the peritoneal cavity and the blood plasma. Glucose and bicarbonate (and/ or lactate), which are usually present in high concentrations in the dialysate move to the plasma. Other solutes such as potassium, phosphate and toxins move from the plasma to the peritoneal cavity. The second is "convection" due to the water flow ("ultrafiltration") that drags solutes across the membrane (from the plasma to the peritoneal cavity). There have been many ways of describing solute flow in the literature since small solute clearance was the main priority of many early PD models. In the following sections, we discuss how the solute flows can be calculated depending on the complexity required. In subsections, we also discuss how to obtain the parameters for the solute flow equations.

The Patlak equation captures such a two-part transport of solutes across the thick inhomogeneous peritoneal membrane (see equation 4.1, Figure 5) [88]. The Patlak equation is essentially an extension of the second equation proposed by Kedem and Katchalsky's thermodynamics-driven volume and solute flux across a thin semi-permeable membrane [48]. First, there is a term that models the diffusive flux across the membrane due to the concentration gradient, given by Fick's first law. The second part is the convective transfer arising from ultrafiltration.

$$J_s = -DA\frac{dC}{dx} + J_v(1-\sigma)C \qquad 4.1$$

where $D$ is the diffusion coefficient ($\mathrm{m^2 s^{-1}}$) and $A$ is the effective surface area available for diffusion ($\mathrm{m^2}$), $C$ is the intramembrane concentration (mol/l) and $J_\mathrm{v}$ is the volume flow ($\mathrm{l\ s^{-1}}$).

Rearranging and integrating the ordinary differential equation 4.1 gives

$$J_s = J_v(1-\sigma)\frac{c_p - c_D e^{-Pe}}{1-e^{-Pe}}$$

$$Pe = \frac{J_\mathrm{v}(1-\sigma)}{MTAC} \qquad 4.2$$

where $c_\mathrm{p}$ and $c_\mathrm{D}$ are the plasma and dialysate solute concentration respectively, $Pe$ is the Péclet number which is a ratio of diffusional and convectional mass transfer (dimensionless), $MTAC$ is the diffusion capacity (“mass transfer area coefficient”) ($\mathrm{l\ s^{-1}}$), $\sigma$ is the reflection coefficient (dimensionless) and $J_\mathrm{v}$ is the volume flow ($\mathrm{l\ s^{-1}}$). For details of the derivation, we refer the reader to [65].

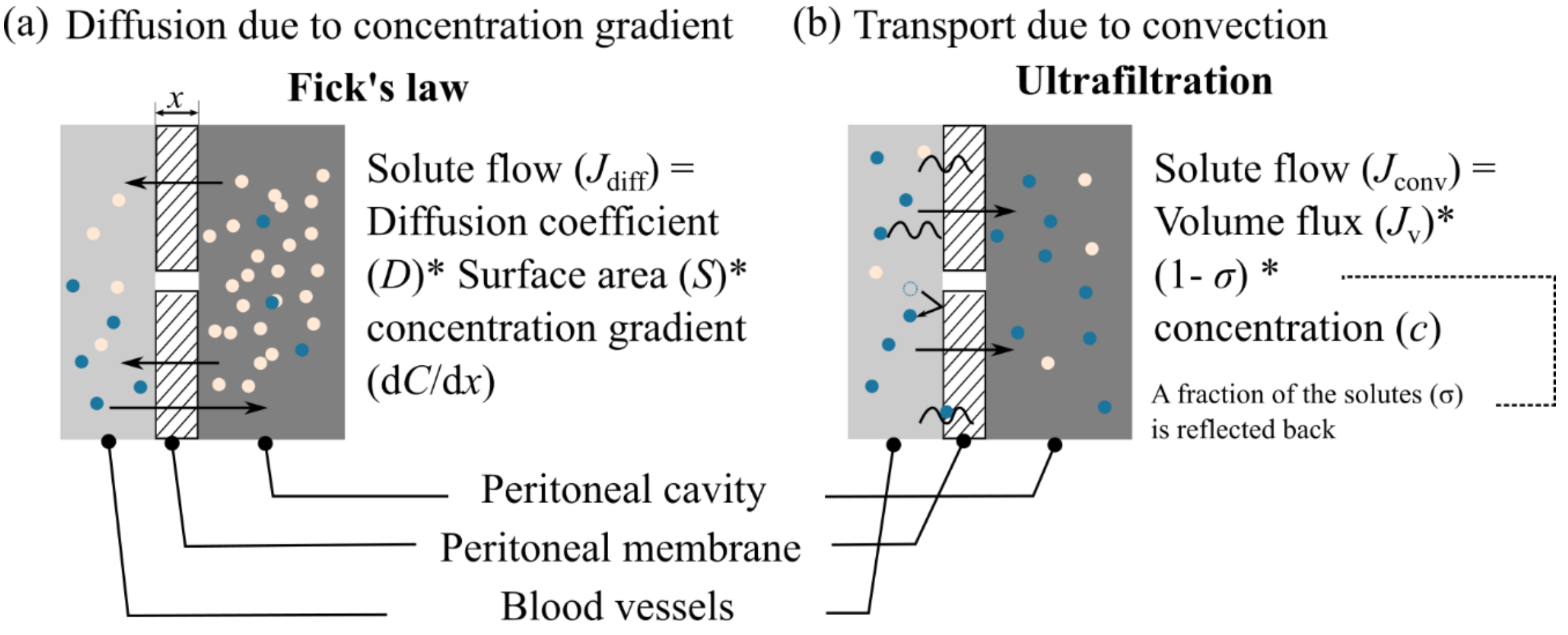

**Figure 5:** Schematic depiction of the solute flow due to diffusion and convection over the peritoneal membrane.

There are other representations of diffusion and convection in various models. Pure diffusion models are sometimes used to define solute transport such as Leypoldt *et al.* and Villarroel *et al.* [36, 89, 90] (equation **Error! Reference source not found.**),

$$V\frac{dc_\mathrm{D}}{dt} = MTAC\ (c_\mathrm{p} - c_\mathrm{D}) \qquad 4.3$$

where $V$ is the peritoneal dialysate volume.

Babb *et al.* use the following equation to represent the solute flux [52, 75] (equation 4.4),

$$V\frac{dc_D}{dt} = MTAC\left(c_{p_0} - c_D\right) + SiCo \cdot \boldsymbol{c} \cdot UFR \qquad 4.4$$

where $SiCo$ is the sieving or transmittance coefficient or membrane selectivity of the particular solute and $UFR$ is the ultrafiltration rate (ml/min, for calculation of parameter, see section 3.1.6). Different interpretations of $\boldsymbol{c}$ (mol/ml) are reported in literature:

| | |
|---|---|
| $\boldsymbol{c} = c_p$ | *when* UF *is considered from only one direction [52, 91] or* UF *is large [75]* |
| $\boldsymbol{c} = (1-f)c_p + fc_D$ | *when it is considered as the intramembrane solute concentration [34, 76, 85]* |
| $\boldsymbol{c} = \frac{c_p + c_D}{2}$ | *when* UF *is low [75]* |

Here, $f$ is a function of the Péclet number, which is by itself a ratio of convection to diffusion, given by

$$f = \frac{1}{Pe} - \frac{1}{e^{Pe} - 1} \qquad 4.5$$

Graff and Fugleberg *et al.* compared six models of the peritoneal solute transport of urea, creatinine, glucose, potassium and phosphate [77-82], using equation 4.6

$$V\frac{dc_{\mathrm{D}}}{dt} = \left(\underbrace{MTAC\big(fct \cdot c_{\mathrm{p}} - c_{\mathrm{D}}\big)}_{diffusive} + \underbrace{SiCo \cdot UF \cdot \boldsymbol{c}}_{Non-lymphatic} - \underbrace{L \cdot C^{*}}_{lymphatic}\right) \qquad 4.6$$

where $fct$ is the equilibrium ratio for solute concentration in dialysate and plasma concentration ($\frac{c_{\mathrm{D}}}{c_{\mathrm{p}}}$), $SiCo$ is the sieving coefficient, $\boldsymbol{c}$ is the intramembrane solute concentration to account for the non-lymphatic convective transfer across the peritoneal membrane and $C^{*}$ is the concentration in the lymph vessels which depends on the direction of the lymphatic flow ($C^{*} = c_{\mathrm{D}}$ if flow is from peritoneal cavity to lymph (vessels) or else $C^{*} = c_{\mathrm{p}}$). They found that glucose transport is purely diffusive while lymphatic flow was important in urea and creatinine transport. Non-lymphatic convective transport is important for urea, creatinine, potassium, phosphate and sodium with sieving coefficient close to 1 (all molecules passing through). Their models showcase that one general model cannot be applied across all body solutes but differs depending on the size, concentration and diffusivity of solutes.

The lymphatic flow is denoted by $L$. Other studies that have included lymphatics flow are Waniewski et al. [92], who show that MTAC values are underestimated for total protein if lymphatic flow is neglected (equation 4.7).

$$V\frac{dc_D}{dt} = \left(MTAC\left(c_p - c_D\right) + (UFR + L)\left[(1-f)c_p + fc_D\right] - Lc_D\right) \qquad 4.7$$

The fitting constant $f$ is usually to be fitted for all solutes separately but an assumption of 0.5 works for most small solutes excluding sodium [83].

Multi-compartment models such as Flessner et al. [33, 93], consider the mass transfer from the capillaries into the surrounding peritoneal tissue and lymphatics to be the contributors for solute flux in the peritoneal cavity (equation 4.8).

$$V\frac{dc_D}{dt} = \left(\underbrace{af \cdot S \cdot \frac{D_t}{\tau} \cdot \left.\frac{\partial c_t}{\partial x}\right|_{x=0}}_{diffusion} - \underbrace{rJ'_v|_{x=x}\,(af \cdot S)c_t|_{x=0} - J'_v|_{x=0}(af \cdot S)c_D}_{convection}\right) \qquad 4.8$$

where $c_t$ is the concentration of the solute in the surrounding tissue, $S$ is the surface area of the peritoneal membrane ($\mathrm{m}^2$) and $D_t$ is the diffusion coefficient in the tissue ($\mathrm{m}^2\mathrm{s}^{-1}$), $\tau$ is the tortuosity (dimensionless) and $r$ is the retardation factor (dimensionless) and $J'_v$ is the local volume flux at distance $x$ into the tissue ($\mathrm{l\ m}^{-2}\mathrm{s}^{-1}$),

$$J'_{\mathrm{v}}|_{x=x} = (J_{\mathrm{v}}ax_t - J_{\mathrm{v}}ax)/(af \cdot S) \qquad 4.8a$$

where $a$ is the capillary surface area per unit tissue ($\mathrm{m}^{-1}$) and $x_t$ is the thickness of the tissue (m).

Gotch added the drain flow rate ($= J_{fill} + UFR$) to derive the solute flux into the peritoneal cavity for single pass CFPD as [87] (equation 4.9),

$$V\frac{\mathrm{d}c_D}{\mathrm{d}t} = \left(MTAC\left(c_{\mathrm{p}} - c_{\mathrm{D}}\right) + UF \cdot \; SiCo \cdot \left(0.67c_{\mathrm{p}} + 0.33c_{\mathrm{D}}\right) - (J_{\mathrm{fill}} + UFR)c_{\mathrm{D}}\right) \qquad 4.9$$

Öberg et al. used the three-pore model for both volume and solute flux for CFPD [32]. Volume fluxes are represented in equation 3.3. Using the volume flux and Patlak equation (equation 4.2), they calculated the solute fluxes as equation 4.10,

$$V\frac{\mathrm{d}c_{\mathrm{D}}}{\mathrm{d}t} = J_{\mathrm{sS}} + J_{\mathrm{sL}} - c_D(J_{\mathrm{vC}} + J_{\mathrm{vS}} + J_{\mathrm{vL}} + J_{\mathrm{fill}}) + c_{\mathrm{D0}}J_{\mathrm{fill}} \qquad 4.10$$

where $J_{s\mathrm{S}}$ and $J_{\mathrm{sL}}$ are the solute fluxes over the small and large pores (solute flux over the ultrasmall pores is non-existent). From equation 4.2,

$$J_{sS} = J_{vS}(1-\sigma)\frac{c_p - c_D e^{-Pe,S}}{1 - e^{-Pe,S}}$$

$$J_{sL} = J_{vL}(1-\sigma)\frac{c_p - c_D e^{-Pe,L}}{1 - e^{-Pe,L}}$$

where $Pe$ is calculated from equation 4.2.

$J_{\mathrm{vC}}$, $J_{\mathrm{vS}}$, and $J_{\mathrm{vL}}$ are the volume fluxes over the three types of pores, $J_{\mathrm{fill}}$ is the fill flow rate and $c_{\mathrm{D0}}$ is the solute concentration in the fresh dialysate.

## 4.1 HOW TO CALCULATE THE SOLUTE FLOW PARAMETERS

After calculating the volume flow, one can use it to calculate the solute flux. Standard peritoneal permeability analysis (SPA) is a standardised tool used to assess the membrane transport properties of a specific patient. These assessments are then used to determine the PD prescription for the specific patient [94]. For a SPA test a static dwell is used with regular dialysate sampling and blood sampling [95].

### 4.1.1 Diffusion Coefficient, $D$

The diffusion coefficient, $D$ ($\mathrm{m^2s^{-1}}$) can be calculated for charged and uncharged particles through liquid flow at low flow rates,

$$D = \frac{\mu \mathrm{k_B} T}{q} \qquad \textit{For charged solutes} \qquad 4.11$$

$$D = \frac{\mathrm{k_B} T}{6\pi \eta r_\mathrm{s}} \qquad \textit{For uncharged solutes}$$

where $\mu$ is the electric mobility of the solute $\left(\frac{\mathrm{m^2}}{\mathrm{Vs}}\right)$, $k_B$ is the Boltzmann constant (Joule, J per Kelvin), $T$ is the temperature (K), $q$ is the charge of the solute (Coulomb), $\eta$ is the dynamic viscosity $(\mathrm{Pa \cdot s})$ and $r_\mathrm{s}$ is the solute radius (m).

### 4.1.2 Mass transfer area coefficients, $MTAC$

The capacity for diffusion MTAC ($\mathrm{ml\ min^{-1}}$) is the maximal diffusive clearance and can be calculated from

$$MTAC = D\frac{A_0}{\Delta x}\frac{A}{A_0}$$

where $A_0/\Delta x$ is the unrestricted surface area to diffusion length ratio (in cm). Typically a patient with an average peritoneal solute transfer rate has an $A_0/\Delta x$ of 25,000 cm, whereas a value < 16,000 cm or > 40,000 cm may indicate slow- and fast peritoneal transport, respectively. The factor $A/A_0$ represents the diffusive hindrance factor and is usually estimated using the equation by Mason, Wendt and Bresler [62], as follows

$$\frac{A}{A_0} = \frac{(1-\lambda)^{9/2}}{1-\ 0.3956\lambda +\ 1.0616\lambda^2}$$

where $\lambda$ is the solute to membrane pore radius ratio. MTAC may also be estimated from experimental data [83, 91, 96]

$$MTAC = \frac{V_t}{t} \ln \frac{V_0^{1-f}(c_p - c_{D,0})}{V_t^{1-f}(c_p - c_{D,t})} \qquad 4.12$$

where $V_t$ and $V_0$ is the intraperitoneal volume at time $t$ and 0, $f$ is from equation 4.5 and $c_{D,0}$ and $c_{D,t}$ is the dialysate solute concentration at time 0 and t respectively and $c_p$ is the plasma solute concentration.

Keshaviah *et al.* compared different functions for MTAC and found that the parabolic and negative exponential functions for urea, creatinine and glucose best fit the dialysate volume profile [68],

$$MTAC = a_1 + a_2V + a_3V^2$$
$$MTAC = a_1[1.0 - \exp(a_2V)] \qquad 4.13$$

where $a_1, a_2, a_3$ are fitting constants that are different for each of the three solutes.

### 4.1.3 Sieving coefficients, $SiCo$

The sieving coefficient, $SiCo$ (dimensionless) is usually fitted in a model [52, 77-82], microscopy study [97] or calculated from experimental observations such as the formula derived by Chen et al. [59],

$$SiCo = \frac{1}{c_{\mathrm{p}}}\left(\frac{c_{\mathrm{D},t}\,V_t + c_{\mathrm{D},g}\,Cl_{\mathrm{albumin}} - c_{\mathrm{D},0}V_0}{V_{\mathrm{UF}}}\right) \qquad 4.14$$

where $V_t$ and $V_0$ is the intraperitoneal volume at time $t$ and 0, and $c_{\mathrm{D},0}$ and $c_{\mathrm{D},t}$ is the dialysate solute concentration at time 0 and t respectively and $c_{\mathrm{p}}$ is the plasma solute concentration, $V_{\mathrm{UF}}$ is the net ultrafiltration volume and $Cl_{\mathrm{albumin}}$ is the clearance of albumin.

Rippe et al. also calculated the sieving coefficient incorporating hematocrit [98],

$$SiCo = 1 - \left(\frac{c_t}{c^*}\left\{1 - \left[\frac{(1 - H_0)(1 - c_0/c_t)}{1 - H_0/H_t}\right]\right\}\right) \qquad 4.15$$

where $H_0 . H_t$ are the initial and final hematocrit and $c_0, c_t$ and $c^*$ are the initial, final and average plasma proteins concentrations.

It can also be simply calculated from the reflection coefficients calculated in section 3.1.2,

$$SiCo = 1 - \sigma \qquad 4.16$$

# 5 Summary of models for peritoneal dialysis

In this section, we provide a comprehensive look at the different models of PD that have been developed throughout the years (1966-2019). We looked for publications that mentioned "kinetic modelling", "mathematical model", "computational model" with "peritoneal dialysis". In total, we found 18 distinct models of PD. We have mentioned what kind of compartments were taken into consideration along with what was the main aim/hypotheses of the modelling effort. Modelling was used as a tool to establish the efficiency of a PD model (5 out of 18) [32, 35, 69, 85, 99] while four models were designed to understand the fundamentals of solute and volume transport [36, 52, 89, 100]. We have also characterised what types of transport processes were included in the volume and solute transport equations of the model. Three of the 18 models were purely diffusive [35, 36, 99] while other models (13 out of 18) included convection and diffusion into the surrounding tissue [74, 101, 102]. Some models were generalised for any type of solute [32, 75, 102] while some were prepared for specific solutes only [77-82, 90]. Table 1 shows an overview of the various PD models that have been published throughout the years. Depending on the solute(s) or drug and which part of the PD process one is interested in, one can choose any of these models.

**Table 1.** Overview of various PD models with different interpretations of volume and solute flux, types of compartments, and PD mode. Some models were developed specifically for a particular type of solute while others can be generalised to other solutes.

| **Model** | Purpose | Type of PD | Method | Volume flux | Solute flux | Solute included | Membrane/compartment | Model developed | Reference |
|---|---|---|---|---|---|---|---|---|---|
| **Kallen (1966)** | To determine PD efficacy in different body sizes by modelling | Static dwell | Modelling | Osmotic gradient | Diffusion | Urea | Body, Peritoneal cavity | Nomogram | [35] |
| **Miller et al (1966)** | To determine the most efficient mode of APD through clinical studies | Intermittent, intermittent recirculating, continuous, continuous recirculating, rapid intermittent, continuous compound dialysis | Modelling + Clinical (n =14) | - | Diffusion | Urea, Creatinine, Uric acid | Body, Peritoneal cavity | Algebraic | [99] |
| **Henderson et al. (1969)** | To make a mathematical model of peritoneal transport by diffusion | Static dwell | Modelling + Clinical (n = 6) | - | Diffusion | Inulin, Urea | Blood, peritoneal cavity | ODE[b] | [36] |
| **Babb et al. (1973)** | To develop a bi-directional mass transfer model for solutes | Static dwell | Modelling + Clinical (n = 3) | Diffusion, Convection | Diffusion, Convection | Urea, Creatinine, Uric acid, Sucrose, B12, Inulin | Capillary blood, peritoneal cavity | ODE | [52] |
| **Villarroel et al. (1977)** | Characterising clearances for different types of PD with modelling | Intermittent, CFPD | Modelling | 0 (intermittent), $J_{outflow}$ (continuous) | Diffusion (intermittent), Diffusion + convection (continuous) | Urea | Blood, peritoneal cavity | Algebraic | [90] |
| **Randerson et al. (1980)** | To develop a model for CAPD | CAPD | Modelling + Clinical (n = 15) | Diffusion, Convection, Residual renal clearance | Diffusion, Convection, Metabolic generation, Residual renal clearance | Urea, Creatinine, B12 | Body + Peritoneal cavity (Urea, Creatinine), Extracellular compartment + intracellular compartment + Peritoneal cavity (B12) | ODE | [85] |
| **Garred et al. (1983)** | To develop a model for mass transfer in CAPD | CAPD | Modelling | Diffusion, Convection | Diffusion, Convection | Urea, Creatinine, B12 | Blood, Peritoneal cavity | ODE | [91] |
| **Flessner et al. (1985)** | To develop a model of peritoneal transport that includes tissue diffusion and convection. | Static dwell | Modelling | Diffusion convection, lymphatic absorption | Diffusion, Convection | Sucrose | Peritoneal cavity, Peritoneal tissue, Distribution volume, body exchange compartment | PDE[c] | [33] |
| **Krediet et al. (1986)** | To determine MTAC[d] for different solutes by a first order kinetic model of solute mass transfer. | CAPD | Modelling + Clinical (n = 11) | Diffusion, convection | Diffusion, convection | Urea, Lactate, Creatinine, Glucose, Kanamycin, Inulin | Blood, Peritoneal cavity | ODE | [96] |
| **Jaffrin et al. (1987)** | To determine the variation in CAPD using a one-pool model varying dwell time and glucose concentration. | CAPD | Modelling | Osmotic gradient | Diffusion, convection | Solute independent | Blood, peritoneal cavity | ODE | [75] |
| **Mactier et al. (1988)** | To determine the contribution of peritoneal cavity lymphatic absorption to the ultrafiltration and solute transfer. | CAPD | Modelling + Clinical (n = 10) | Transcapillary ultrafiltraion, Lymphatic absorption | - | Creatinine, Glucose | - | Algebraic | [100] |
| **Leypoldt et al. (1988)** | To distinguish between indicator dilution volume and true dialysate volume. | Static dwell | Modelling + experiment (rabbit, n = 9) | - | Diffusion, lymphatics | Creatinine | Blood, lymphatics + Peritoneal tissue | ODE | [89] |
| **Vonesh et al. (1991)** | To use modelling as a predictive tool for suggesting PD type to patients. | CCPD, Tidal PD | Modelling + Clinical (n = 5, different PD types) | Diffusion, convection | Diffusion, convection | Urea, Creatinine, Glucose, $\beta$microglobulin | Body, Peritoneal cavity | ODE | [76] |
| **Waniewski, Werynski et al. (1991)** | To simplify Garred model [91] for small solute transport | Static dwell | Modelling + clinical studies (n = 21) | Diffusion, Convection | Diffusion, Convection | Urea, Creatinine, Glucose, Potassium, Sodium, Protein | Blood, Peritoneal cavity | Algebraic | [83] |
| **Rippe (1991)** | To develop a new model for CAPD assuming the peritoneal membrane is mainly composed of three types of pores. | CAPD | Modelling | Ultrafiltration, lymphatics | Diffusion, convection | Glucose, Urea, Sodium, albumin, Phosphate, $\beta$microglobulin | Body, Peritoneal Cavity | ODE | [42] |
| **Graff and Fugleberg (1994)** | To determine the best solute transport mechanism for different solutes. | Static dwell | Modelling + Clinical (n = 21 to 26) | Ultrafiltration, Lymphatics | Diffusion, Convection (non-lymphatic and lymphatic) | Urea, Glucose, Phosphate, Creatinine, Potassium, Sodium | Body, Peritoneal Cavity | ODE | [77-82] |
| **Gotch (2002)** | To develop a kinetic model of CFPD. | Single pass CFPD | Modeling | Ultrafiltration | Diffusion, Convection | Urea | Body, Peritoneal cavity, (External Dialyser) | ODE, Empirical | [87] |
| **Akonur et al. (2010)** | To use TPM for optimisation of drain phase in static dwell | Static dwell | Modelling | Based on TPM [66] with new biphasic equation for drain | Based on TPM [66] | Based on TPM [66] | Based on TPM [66] | ODE | [103] |
| **Akonur et al. (2015)** | To modify TPM to include α-amylase activity in icodextrin kinetics | Static dwell | Modelling | Based on TPM [66] | Based on TPM [66] + first order degradation of higher weight fractions of icodextrin | Based on TPM [66] | Based on TPM [66] | ODE | [104] |
| **Oberg et al. (2017)** | To extending classic TPM to include the fill and drain phases of dwell. | APD | Modelling | Based on TPM [66] | Based on TPM [66] | Based on TPM [66] | Based on TPM [66] | ODE | [50] |
| **Oberg et al (2019)** | To extend TPM for CFPD and determine the ultrafiltration rate. | CFPD | Modelling | Based on TPM [66] | Based on TPM [66] | Based on TPM [66] | Based on TPM [66] | ODE | [32] |
| **Lee et al. (2020)** | To model steady concentration PD (SCPD) | SCPD | Modelling | Based on TPM [66] | Based on TPM [66] + Infusion rate of glucose | Based on TPM [66] | Based on TPM [66] | ODE | [105] |
| **Wolf et al. (2021)** | To extend TPM to determine acid-base kinetics during PD | Static dwell | Modelling | Based on TPM [66] | Based on TPM [66] + $CO_2$ conversion to bicarbonate | Same as [66] – Creatinine - $\beta$ microglobulin +Lactate + Bicarbonate +Calcium + Magnesium | Based on TPM [66] | ODE | [106] |
| **Stachowska-Pietka et al. (2023)** | To modify TPM to include α-amylase activity in icodextrin kinetics | Static dwell | Modelling + Clinical (*n* = 11) | Based on TPM [66] | Based on TPM [66] + first order kinetics of icodextrin hydrolysis including α-amylase concentration | Based on TPM [66] | Based on TPM [66] | ODE | [107] |
| **Hartinger et al. (2023)** | To make a population pharmacokinetic model of vancomycin | Static dwell | Modelling + Clinical (*n* = 41) | - | First order clearance | Vancomycin | Peritoneal cavity, one and two-compartment models for rest of the body | PK[e] | [108] |

[a] particularly for small solutes such as urea

[b] Ordinary differential equations: We assume that the compartment is homogeneously mixed.

[c] Partial differential equations: Distribution within the compartment is important.

[d] Mass transfer are area coefficients

[e] Pharmacokinetic model: Series of ODE to describe the rates of drug absorption, distribution, metabolism and elimination to predict drug concentration changes over time

# 6 TEST CASE FOR DEVELOPING A MATHEMATICAL MODEL OF PD

Many PD-related questions could be answered with mathematical modelling, for example:

- What is the optimal glucose concentration for PD, balancing ultrafiltration (volume and efficiency) versus adverse effects (e.g. increased peritonitis risk)?
- What is the influence of two catheters *vs* one single lumen catheter on solute clearance in continuous flow *vs* tidal PD?
- How does the intraperitoneal dialysate volume affect solute clearance in CAPD or CFPD?
- Which flow rates are ideal in CFPD?
- What is the clearance of a particular solute/drug for a particular PD modality (CAPD, APD or CFPD)?
- How does tidal PD with partial drainage of the solution, leaving a residual volume in the peritoneal cavity, affect solute clearance as compared to complete drainage?

In Table 2, we demonstrate how we can create and implement our mathematical model if we know which research question we want to answer, what experimental data or literature data we possess and what kind of maths is necessary for the model. With the realised model, we can then play with the parameters to determine an optimal treatment scenario for the patient as we can analyse both the short-term and long-term effects of a session.

**Table 2:** Steps to design a mathematical model from scratch. An example scenario explains how with the objective and preliminary data specified, we can further build an existing model to answer a particular PD-related question.

| | | Steps to creating a mathematical model | *Example scenario* |
|---|---|---|---|
| 1 | INPUT | Have the problem clearly specified | What is the clearance of drug X is given to the patient via PD? What is the influence of drug X on solute clearance? |
| 2 | | What data do I have? | The concentration of solutes ($c_D$) and drug ($c_{drug}$)in the fluid, device flow rates ($J_{fill}$, $J_{drain}$), distribution volume and intra-peritoneal volume, diffusion capacities of the solutes and drug ($MTAC$). $MTAC$ values can be obtained from studies like [109], where they analyse the transport of model compounds according to weight, acidity, partition coefficient ($\Phi$, see section 3.1.2). Note that an increase in MTAC should be taken into account for CFPD (several-fold increase may occur) |
| 3 | | Do I consider lymphatics to be involved in removal of this drug? | Depending on the answer, one can put the lymphatic flow rate, $L$ = 0 for the ones that don't have lymphatics involved at all. Certain drugs might inhibit solute |
| | | | transport to blood but encourage lymphatic transport [110, 111]. |
| 4 | | Is the drug removed primarily by diffusion (like glucose) or also via convection? | There are many studies for drug clearance by peritoneal membrane, especially chemotherapy drugs [112, 113]. Dedrick *et al.* designed a pharmacokinetic model of drug clearance by the peritoneal membrane to select the one better suited for a clinical trial [114]. Having an idea of the removal kinetics would help to choose between a diffusion only model [35, 99] or convection+diffusion model [32, 87]. |
| 5 | | Is the solute clearance inhibited by the drug X? | Studies show that drugs like furosemide or ACE inhibitors might inhibit solute transport across the peritoneal membrane [115, 116]. This can be modelled by a reduced $MTAC$ for the solute in question. |
| 6 | | Does the drug penetrate the surrounding tissue? | Dedrick *et al.* have shown in their model that for slowly reacting drugs there can be surface penetration upto $(D/k)^{0.5}$ where $D$ is the diffusivity of the drug and $k$ is the rate constrant of drug removal from tissue [117]. If significant, it might be necessary to use the Flessner model [102]. |
| 7 | | Is it a static dwell or continuous flow model? | Static dwell model such as [80, 83, 89] or continuous flow model like [69, 87, 96] could be chosen. |
| 8 | | Do I have time series data of the dialysate solute concentration? | If yes, the model that is chosen in the previous steps 3-7 could be fitted to obtain the correct $MTAC$ value or other previously assumed parameters. If no, with the PD effluent, one can estimate the average $MTAC$ during a particular session. Note that, in both cases, average $MTAC$ is estimated as most often the parameter reduces during the dwell time [68]. |
| 9 | OUTPUT | Can I calculate clearance? | With a precise parameter set, clearance of drug X and other solutes could be calculated in any time range for a specific patient. |

## 7 DISCUSSION

In this review, we give an overview of the physical principles that govern peritoneal dialysis and summarize the essential (differential) equations for volume and solute flux that are required to model peritoneal dialysis. These models are sometimes based on simple principles and parameters are lumped together to study a compartmentalised version of the body [35, 52, 77-82], while other models are very complex and require many parameters to capture the physical processes in depth [32, 93, 102]. We also list the different ways the various simulation parameters are obtained. Most of the times, the parameters are derived directly from the patient data (effective peritoneal surface area, dialysate clearance etc), some of the parameters are derived from analytical formulae (reflection coefficient, hydrostatic and osmotic pressure difference), others are derived from the model itself (ultrafiltration, pore contribution to ultrafiltration, fractional hydraulic conductance) while some parameters may be fitted to obtain the best (patient-specific) interpretation of the dialysate and plasma solute concentration. Each model can be modified to fit the required patient data and get outputs as ultrafiltration rate, mass transfer area coefficients and residual volumes.

The human body is a complex system, which makes it difficult to model. Modelling its various parts will eventually lead us to understand the whole. PD models have generally developed in the direction of including more transport processes. The first PD models were purely diffusive which is representative of small molecular transport such as that of urea and creatinine [35, 36, 99]. As the importance of mid-sized molecules were better understood, convection was added to the models as this is an important route of elimination for these compounds [52, 90]. Lymphatic absorption [32, 87] and peritoneal tissue surface area [102] has since been added to the models. As per the new ISPD guidelines, ultrafiltration is a crucial parameter to determine the efficacy of a particular PD treatment [118]. Thus we see that recent models are also trying to model time-dependent ultrafiltration in patients [32]. This requires rigorous evaluations and comparisons of ultrafiltration rates. In section 3.1.6, we discuss multiple ways to obtain ultrafiltration rate and which efforts can be made to make the best estimate.

With an increase in computational power and physical understanding, the model complexity can be increased to create a better picture of the underlying mechanisms in the patient's peritoneal cavity. Herein lies a caveat. The increased complexity makes the model space hyper-parameterised. Finding efficient parameter combinations which satisfy physiological relations is difficult to achieve without multiple assumptions. This can be achieved by fitting parameters to clinical data (for example, in [119], Stachowska-Pietka et al. fitted 16 parameters while assuming 9 parameters to be fixed) and calibrating the model in a different scenario (they showed that the predictions of interstitial concentration of mannitol in rat abdominal wall were in agreement with experiments[119]). Future models can be improved to include metabolism of glucose in the peritoneum and how it affects the peritoneal membrane, the positioning of the catheters in specific patients and cellular contributions to the solute clearances (for example for glucose [120]). Future PD models can also be made spatial (using for example partial differential equations) to understand the spatial influence of the flow and solute gradient. How does the continuous flow change the boundary layer of the peritoneal membrane? What happens in the residual volume? What fraction of recirculation occurs just at the tip of the catheter? Can patient-reported outcome measures be linked to PD (efficacy) parameters? Partial differential equation modelling (used for space localisation) of the peritoneal cavity may help us answer these questions in the future. Agent based modelling (ABM) can also be helpful in this aspect. ABM models take into account different agents (e.g. patient characteristics, residual kidney function, smoking, dietary habits, transport status, sex, weight, osmotic agent etc.) to assess the problem and make decisions based on a complex behaviour pattern. ABM models have already been used to determine the optimum treatment pathway for HD patients based on patients', nephrologists' and surgeons' attributes [121]. From recent developments in PD, we definitely see that patient membrane characteristics and preferences play a huge role in managing the patient. A lot of focus has to be dedicated to make significant changes to the present computational models to make them personalised. Efforts have already been made in the past to suggest the best mode of PD treatment for different patient characteristics [32, 76, 90, 99, 122] but more work needs to be done on personalising a single PD mode based on the sex, age, weight, peritoneal characteristics and residual kidney function of the patient. Still, there are many aspects that a virtual PD system cannot cover, for example catheter dysfunction and medication nonadherence.

We can picture the growth of mathematical modelling of PD in two complementary directions: fundamental understanding and personalisation. More technical components can be introduced to the existing models and parameters can be analysed to facilitate a faster understanding of the new devices and optimise them before market entry. One aspect that has been little explored is to employ computational modelling to evaluate different bio-compatible osmotic agents replacing glucose to avoid the adverse effects of glucose and glucose by-products in PD patients. Recent advances in modelling includes using PD models to create virtual clinical trials and personalise treatments specific to the patient with the help of nephrologists.

*Virtual Patients*

The interest in virtual patients is growing as we have seen with PBPK (physiologically based pharmacokinetic) and PKPD (pharmacokinetic and pharmacodynamic) modelling studies being mandatory in drug studies [123-125]. Similarly, *in silico* modelling could become essential to medical device development. To realise this, we need digital patient twins. Enabling the use of a virtual platform to test and develop medical devices, would reduce risk and regulatory burden. The FDA also has directed funds towards the establishment of computational modelling as a regulatory tool [126]. A growing virtual patient database (including animals) could mean that with a computational model, we could identify the risk groups and drug side effects, reduce animal testing and establish safe protocols.

*Complexity of PD devices*

Currently, there are several CFPD devices under development using continuous sorbent-based dialysate regeneration (preclinical or clinical developmental phase) [127, 128]. To realise these devices in mathematical terms, future work needs to focus on adding flow through a sorbent chamber. There needs to be an evaluation of dead volume- and recirculation-related loss of efficiency in the system due to application of rapid flow cycling via a single lumen catheter. Considering all the potential parameter settings, which likely influence each other, investigating all parameter combinations is very time consuming requiring large clinical trials, with many resources and tremendous planning [129]. Computational modelling can help in rapid optimisation and avoid unnecessary scenario testing through careful calibration and validation.

*Personalisation*

Other problems with clinical trials include lack of randomisation of patients (inclusivity in terms of sex, age and gender etc.), difficulty in studying long term effects (due to patient drop out (e.g. due to kidney transplantation)), lack of blinding, the learning curve (an already existing system may be easier to handle) and general reluctance (from doctors and patients) to try new technologies [25]. Chronic kidney disease patients also use various types of medications which may interfere with the efficiency of the PD session. With mathematical modelling, we can conduct virtual PD trials with an inclusive patient spread (age, sex, stage of renal disease, transport parameters, intra-abdominal volume etc.) and study the short- and long-term effects of different modes of PD. PD prescription models such as PatientOnLine [130, 131] and PD ADEQUEST [132, 133] have been developed to optimize and personalize PD prescription which have also been validated in multicentre studies. However, focus of these prescription models is on optimization of dialysis dose and ultrafiltration and not on patient quality of life. Incorporation of more patient specific factors [134-136] into the model may further personalize PD prescription and contribute to patient well-being.

# 8 CONCLUSION

*In silico* modelling is a powerful tool that can be used to understand gaps in knowledge of new and old PD devices and facilitate their quick and efficient transition from *in vitro* to *in vivo* to patients. In this review, we have looked at the different modelling approaches explored throughout the years to model PD and the essential components in a PD compartmental model. We are optimistic that a joint effort of computational modellers and clinicians can help not only in technical improvement of PD technology but also personalisation of PD treatment to ensure a higher quality of life for patients.

# 9 DECLARATIONS

## 9.1 Ethics approval and consent to participate

Not applicable

## 9.2 Consent for publication

Not applicable

## 9.3 Availability of data and materials

Not applicable

## 9.4 Competing interests

The authors declare that they have no competing interests.

## 9.5 Funding

This work is supported by the partners of Regenerative Medicine Crossing Borders (RegMed XB), a public-private partnership that uses regenerative medicine strategies to cure common chronic diseases, by the Dutch Kidney Foundation and Dutch Ministry of Economic Affairs by means of the PPP Allowance made available by the Top Sector Life Sciences & Health to stimulate public-private partnerships (DKF project code PPS08), by a grant from the Dutch Kidney Foundation ( 22OK1018) and by the European Union (CORDIAL, Horizon 2020 research and innovation program, grant agreement no. 945207).

## 9.6 Author's contributions

Sangita Swapnasrita participated in Writing and Original Draft Preparation; Joost C de Vries, Aurélie MF Carlier, Carl M Öberg, Karin GF Gerritsen participated in Writing and Review and Editing; Funding Acquisition was done by Aurélie MF Carlier and Karin GF Gerritsen.

## 9.7 Acknowledgements

Not applicable

# 10 GLOSSARY

A = fitting constant for exponential decrease of UF over time (dimensionless)

$af$ = Fraction of peritoneal membrane in contact with fluid (dimensionless); $af = 16.18 * \frac{1-e^{-0.00077*V}}{13.3187}$ [69]

$c_{\mathrm{D}}$ = dialysate solute concentration (mmol/l or mmol/m$^3$), proteins are displayed in g/L or g/dL.

$c_{\mathrm{drain}}$ = drain solute concentration (mmol/l)

$c_{\mathrm{p}}$ = peripheral vein plasma water solute concentration (mmol/l), proteins are displayed in g/L or g/dL.

$\boldsymbol{c}$ = intramembrane solute concentration (mmol/l)

$D$ = diffusion coefficient (dimensionless)

$f$ = function of Pe (dimensionless)

$fct$ = equilibrium $c_p/c_D$ (dimensionless)

$J_{\mathrm{drain}}$ = drain flow rate (l/min)

$J_{\mathrm{fill}}$ = fill flow rate (l/min)

$J_{\mathrm{v}}$ = volume flux (l/min)

$J_{\mathrm{s}}$ = solute flux (mmol/min)

$K$ = permeability coefficient (dimensionless)

$L$ = lymphatic flow (l/min)

$L_p$ = hydraulic conductivity (l/(min.cm$^2$.mmHg))

$MTAC$ = mass transfer area coefficients (ml/min or m$^3$/min)

$P$ = hydrostatic pressure (mmHg)

$Pe$ = Peclet number (dimensionless) to determine the importance of convection over diffusion

$r$ = retardation factor (dimensionless)

$S$ = peritoneal surface area (m$^2$)

$SiCo$ = sieving coefficient (dimensionless)

$t$ = time of session (hr)

$UFR$ = ultrafiltration rate (l/min)

$V$ = intraperitoneal volume (l)

$V_r$ = residual volume (l)

$V_{fill}$= initial fill volume (l)

$V_{drain}$ = drain volume (l)

$W$ = body weight (kg)

**Greek symbols**

$\alpha$ = contribution to ultrafiltration coefficient (dimensionless); for ultrasmall pores $\alpha_C$ = 0.02, small pores $\alpha_S$ = 0.9 and for large pores $\alpha_L$= 0.08

$\beta$ = time constant for exponential decrease of UF over time (dimensionless)

$\lambda$ = solute radius/membrane pore radius (dimensionless)

$\sigma$ = reflection coefficient (dimensionless); calculated for different solutes.

$\pi$ = oncotic pressure (mmHg)

$\Phi$ = equilibrium partition coefficient (dimensionless)

$\tau$ = tortuosity factor (dimensionless)

**Subscripts**

$glu$ = glucose

$t$ = at time t

$fill$ = filling the peritoneal cavity at time 0 (static dwell) or during the session (CFPD, APD)

$drain$ = drain of the peritoneal cavity after (static dwell) or during the dwell (CFPD, APD)

$s$ = solute flux

$v$ = volume flux

$C$ = ultrasmall pores

$S$ = small pores

$L$ = large pores

$r$ = residual

**Acronyms**

APD = automated PD or continuous ambulatory PD

AQP1 = Aquaporin 1

CFPD = Continuous flow PD

ESKD = End stage kidney disease

HD = Haemodialysis

IPV = Intraperitoneal volume

MTAC = Mass transfer area coefficients

OCG = Osmotic conductance to glucose

PD = Peritoneal dialysis

SAPD = Sorbent assisted PD

TPM = Three-pore model